\documentclass[12pt]{article}
\usepackage{pic03}
\usepackage{hyperref}
\usepackage{url}
\usepackage{graphicx}

\begin{document}

\title{ EXPERIMENTAL EVIDENCE FOR HADRON DECONFINEMENT IN $\rm \bar{p}-p$ 
COLLISIONS AT $\sqrt{s} = 1.8$ TeV IN THE FNAL C\O~COLLIDER}

\author{
L. GUTAY ( for the E-735 Collaboration)    \\
{\em Purdue University, Department of Physics} \\
{\em 525 Northwestern Avenue, West Lafayette, IN 47907-2036 USA}}
\maketitle

\baselineskip=14.5pt
\begin{abstract}
We have measured deconfined hadronic volumes, 4.4 $<$ V $<$ 13.0 fm$^3$,
produced by a one-dimensional(1D) expansion. The volumes are directly proportional to the charged particle pseudorapidity densities 
6.75 $<$ dNc/d$\eta <$ 20.2. The hadronization 
temperature is $T= 179.5 \pm 5$(syst) MeV. The hadronization energy density is $\epsilon =1.10 \pm 0.26$ GeV/fm$^3$, 
corresponding to an excitation of 24.8 $\pm$ 6.2 quark-gluon degrees 
of freedom.
\end{abstract}

\baselineskip=17pt

\section*{}
The deconfined
hadronic matter has a threshold at  $ dNc/d\eta $ = 7. 
Hadronization for all events above the threshold 
occurs at a constant energy density and 
temperature. The temperature is constant to 1$\%$ for
6.75 $<$ dNc/d$\eta <$ 20.2.
 If we treat all charged particles as pions then  1.57 $\pm$ 0.25(stat)
pions/$fm^3$ are emitted from the volume. 
We can relate this number of pions/$fm^3$ emitted from the volume V to the 
number of gluons and quarks in the volume (see Fig.1). 
\begin{equation}
\rm n~(pions)~=~n~(gluon) +\frac{n(q)+n(\bar{q})}{2}
\end{equation}
We assume that two flavors of quarks and gluons are involved and that the
number of quark and gluon degrees of freedom scale proportionally.
\begin{equation}
\rm{n(pion)} = V \frac{G(T) 1.202 (kT)^{3}}{\pi^2 \hbar{^3} c^3}
\end{equation}
This corresponds to  $G(T) = 23.5 \pm 6.0$(stat) quark-gluon degrees of freedom.
 Similarly all charged particles contribute to the freeze out energy E .
\begin{equation}
E(freezout) = V  \;
\frac{G(T) \; \pi^{2} \; k^{4}}
     {30 \; \hbar^{3} \; c^{3}}
T^{4} \ . 
\end{equation} 
\begin{figure}
\centerline{\hbox{ \hspace{0.5cm}
\includegraphics[width=8.5cm]{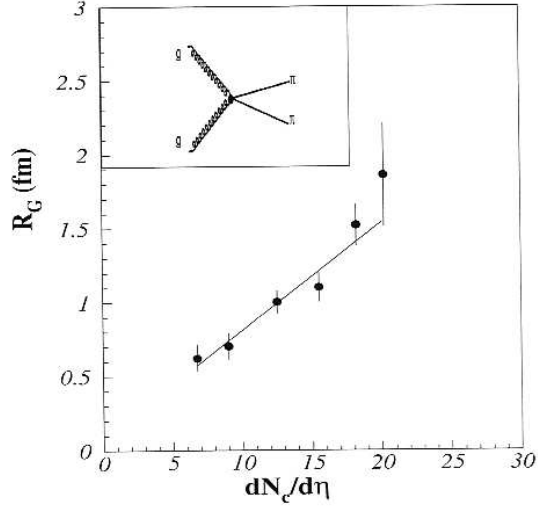}}}
\caption{\it Dependence of the longitudinal Gaussian radius R$_G$ on dN$_C$/d$\eta$.} 
\end{figure}[ht]
The freezout energy density emitted by the volume V is  
$\epsilon =1.10 \pm 0.26(stat)$ GeV/fm$^3$  
and we obtain $G(T) = 24.8 \pm 6.0(stat)$ quark-gluon degrees of freedom.  
In Fig.2. we compare our experimental results with Karsch's lattice gauge 
calculation [2], where  
$\frac{\epsilon}{T^{4}} = \frac{\pi^{2}}{30}~G(T) = 8.15 \pm 2.0$ (stat).
\begin{figure}[ht]
\centerline{\hbox{ \hspace{0.5cm}
\includegraphics[width=8.5cm]{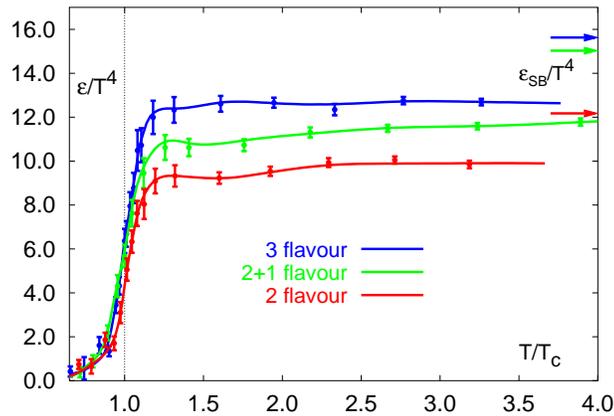}}}
\caption{Lattice gauge calculation of [2]}
\end{figure}


\end{document}